\documentclass[doublecol]{epl2} 

\title{  Equilibration  problem for the generalized Langevin equation}

\author{Abhishek Dhar\inst{1} \and Kshitij Wagh\inst{2} }

\institute{                    
  \inst{1} Raman Research Institute, Bangalore 560080 \\
  \inst{2} Department of Physics and Astronomy, Rutgers University, Piscataway, New Jersey 08854-8019 
}
\pacs{05.40.-a} {Fluctuation phenomena, random processes, noise, and
  Brownian motion} 
\pacs{05.10.Gg} {Stochastic analysis methods}
\pacs{05.70.Ln} {Nonequilibrium and irreversible thermodynamics}

\abstract {We consider the problem of equilibration of a single
oscillator system with dynamics  given by the generalized classical 
Langevin equation.  It is well-known that this dynamics can be obtained if one
considers a model where the single oscillator is coupled to an
infinite bath of harmonic oscillators which are initially in equilibrium. 
Using this equivalence we first determine the conditions  necessary for
equilibration for the case when the system potential is harmonic.
We then give an example with a particular bath where we show that,
even for parameter values where the harmonic case always equilibrates,
with any  finite amount of nonlinearity the system does not 
equilibrate for arbitrary initial conditions. We understand this as a
consequence of the formation of nonlinear localized excitations similar
to the discrete breather modes in nonlinear lattices.   
}

\begin{document}

\maketitle
\def\bea{\begin{eqnarray}}
\def\eea{\end{eqnarray}}
\def\a{\alpha}
\def\d{\delta}
\def\D{\Delta}
\def\p{\partial} 
\def\nn{\nonumber}
\def\r{\rho}
\def\rv{\bar{r}}
\def\la{\langle}
\def\ra{\rangle}
\def\e{\epsilon}
\def\om{\omega}
\def\Om{\Omega}
\def\n{\eta}
\def\g{\gamma}
\def\bFi{{\Phi}}
\def\bM{{ M}}
\def\bY{{\cal{Y}}}
\def\bG{{\Gamma}}
\def\l{\lambda}
\def\break#1{\pagebreak \vspace*{#1}}
\def\f{\frac}
\def\dg{\dagger}
\def\zh{\hat{Z}}
\def\s{\sigma}
\def\bd{{\bf d}}
\def\ba{{\bf a}}
\def\be{{\bf e}}
\def\bb{{\bf b}}
\def\l{\lambda}
\def\tg{\tilde{\gamma}}
\def\tx{\tilde{x}}
\def\teta{\tilde{\eta}}

\section{Introduction}
One of the simplest phenomenological ways of
modeling the interactions of a system 
with a heat bath is through the Langevin equation which, for a single
particle, of unit mass and  moving in a one dimensional potential $V(x)$,
is given by 
\bea
\ddot{x} = -d V(x)/d x-\g \dot{x}+\eta(t)~, 
\label{leq}
\eea 
where  $\eta (t)$ is a Gaussian white
noise which satisfies the fluctuation-dissipation (FD) relation 
$\la \eta(t) \eta(t') \ra = 2 k_B T \g \delta (t-t') $ . Here $T$ is 
the temperature of the heat bath. The dynamics in eq.~(\ref{leq})
ensures that at long times the system reaches thermal
equilibrium. Thus the phase space 
density $P(x,p,t)$, where $p=\dot{x}$, converges in the 
limit $t\to \infty$ to the Boltzmann distribution $e^{-\beta H_s}/Z$ where
$H_s=p^2/2 +V(x)$ and $Z$ is the corresponding partition function.
The proof for this uses the  correspondence between the Langevin
equation and the Fokker-Planck equation \cite{risken}.

However the $\delta$-correlated nature of the noise is 
unphysical and this has led to the study of the generalized Langevin
equation \cite{mori65,kubo66} 
\bea
\ddot{x} = -d V(x)/d x-\int_{-\infty}^t dt' \g (t-t')  \dot{x}(t')
+\eta(t)~, 
\label{gleq}
\eea
where the noise is correlated and the  the dissipative term involves a memory kernel.
The noise is again Gaussian and is related to the dissipative
term  through the generalized FD relation 
\bea
\la \eta(t) \eta(t') \ra = k_B T \g(t-t')~.
\label{gFD}
\eea
 A standard
method of microscopically modeling a heat bath is to consider an
infinite collection of harmonic oscillators, with a distribution of
frequencies, coupled linearly to the system. In that case it can
be shown \cite{zwanzig,weiss99}  that the effective equation of motion of the system is
precisely given by eq.~(\ref{gleq}) where the dissipation kernel
$\g(t)$ depends on the bath oscillator frequencies and the coupling
constants. 
Unlike the case with $\d$-correlated noise there exists no general
proof that, for a general potential $V(x)$, the system will
reach thermal equilibrium at long times.  The reason for this is that
in this case the construction of a Fokker-Planck description is
difficult and is  known in few cases ({\emph{e.g.}} harmonic
oscillator case treated in Ref.~\cite{adelman}).  For the special case of a
harmonic potential and with certain restrictions on the form of
$\g(t)$ one can prove equilibration  by a direct solution
of the equations of motion \cite{kubo}. 
In the quantum mechanical case the oscillator bath model has been
 widely used to model the effects of noise, dissipation and decoherence in
 quantum systems \cite{ford,caldeira}. In this
case the approach to equilibrium has been proved only for a
special class of  potentials for the cases where the system-reservoir
coupling is weak \cite{kac}.

A number of   papers \cite{sroko00,wang,bao} have attempted to understand various aspects 
of the  generalized Langevin equation such as anomalous diffusion,
nonstationarity and ergodicity.  
In this paper we address the question of approach to equilibrium for
the generalized Langevin equation. For a particle in a harmonic
potential $V(x)=\om^2_0 x^2/2$ and 
coupled to a heat bath with a  finite band-width (and hence long-time
memory) we show that  
equilibration is not always ensured and depends on the oscillator
frequency $\om_0$. We find the necessary
conditions for equilibration which is related to the existence of
bound states (localized modes) of the coupled system-plus-bath. 
We note that Ref.~\cite{bao} looks at the
question of ergodicity which is similar to the question addressed
here. One of their results is that the motion of a  harmonic
oscillator coupled to a general heat bath is ergodic.
This is based on the fact  that the coefficients in the 
Fokker-Planck equation  asymptotically approach constant values. Our
work shows that this condition does {\emph{not}} ensure ergodicity.
Next we consider a 
particle in a nonlinear potential $V(x)=\om_0^2 x^2/2 + u x^4/4$ and
coupled to a special bath, the so-called Rubin model \cite{rubin60,weiss99}.
The usual expectation
would be that nonlinearity should help in equilibration. However
surprisingly we find the contrary to be true. For values of $\om_0$ for
which the linear system (with $u=0$) does equilibrate, we show that
switching on  the nonlinearity can lead to loss of equilibration.
We relate this to the formation of nonlinear localized  modes which
are similar to the discrete breather modes found in nonlinear lattices
\cite{sievers,flach}. 

\section{Definition of model}
We consider the following Hamiltonian
of the usual model of a system coupled to a bath of $N$ oscillators:
\bea
H=H_s+ \sum_{\a=1}^N \f{P_\a^2}{2}+\f{\om_\a^2}{2} 
\left( X_\a-\f{c_\a  x}{\om^2_\a} \right)^2 ~, \label{ham}
\eea
where $\{ X_\a,~P_\a \}$ denotes  degrees of
freedom of the $\a^{\rm th}$ bath oscillator ($\a=1,2,...N$), $\om_\a $ is its
frequency,  and $c_\a$ is the strength of the coupling to the
system. For dissipation it is necessary that the frequencies $\om_\a$
have a continuous spectrum in the limit $N \to \infty$.  
We assume that at time $t=t_0$ the system and bath are decoupled
($\{c_\a\}=0$ ).
The bath is in thermal equilibrium and described by the canonical
distribution $e^{-\beta H_b}$ where $H_b=\sum_\a [~P^2_\a/2+\om_\a^2
  X_\a^2/2~]$ and the system is in an arbitrary initial state
$\{x(t_0),~p(t_0)\}$. The coupling is switched on at time $t=t_0$. It
can 
then be shown that eliminating the bath degrees of freedom 
leads to the following equation of motion for the system (for $t > t_0$):
\bea
 \ddot{x} = -\f{d V(x)}{d x}-\g(0) x +\int_{t_0}^t dt' \f{d\g (t-t')}{dt'}  {x}(t')
+\eta(t)~, 
\label{redeq}
\eea
with the dissipation kernel given by
$ \g(t)=\sum_\a ({c_\a^2}/{\om_\a^2}) \cos {(\om_\a t)}$. 
In the limit $t_0 \to -\infty$, using $\g(t \to \infty)=0$, we get eq.~(\ref{gleq}). 
The noise correlations satisfy the generalized FD relation in eq.~(\ref{gFD})
where the noise average $\la ...\ra$ is obtained by averaging 
over the initial conditions of the bath.

\section{Equilibration in a harmonic potential}
We first  
consider the case with $V(x)=\om^2_0 x^2/2$. For this case
eq.~(\ref{redeq}) is a linear inhomogeneous equation whose solution is:
\bea
x(t)&=& H(t-t_0) x(t_0) + G(t-t_0) p(t_0) \nn \\  
&&+ \int_{t_0}^t dt' G(t-t') \eta(t')~, \label{gensol}
\eea
where $H(t)$ and $G(t)$ are the solutions of the
homogeneous part of eq.~(\ref{redeq}) (with $t_0=0$) for initial conditions 
$H(0)=1,~\dot{H}(0)=0$ and  $G(0)=0,~\dot{G}(0)=1$ respectively. 
For equilibration we require that, in the long time limit, the
solution in eq.~(\ref{gensol}) should not
depend on initial conditions. Thus a necessary condition is that
$G(t \to \infty) \to
0$. At large times it can be shown that 
$H(t) \to \dot{G}(t)$ and hence $H(t \to \infty) \to 0$ also.
Let us define the Green's function $G^+(t)= G(t) \theta
(t)$. It is easy to see that $G^+$ satisfies the equation of
motion:
\bea
 \ddot{G^+} + \om_0^2  G^+ +\int_{-\infty}^\infty dt' \g^+ (t-t')
\dot{G}^+(t')= \delta(t)~, \nn 
\eea
where $\g^+(t)=\g(t) \theta(t)$.
Defining the Fourier transforms $G^+(\om)= \int_{-\infty}^\infty dt
e^{i \om t} G^+(t)$ and $\g^+(\om) =\int_{-\infty}^\infty dt e^{i \om t}
\g^+(t)$ we find that $G^+(\om)$ is given by 
\bea
G^+(\om)&=&\f{1}{- (\om^2-\om_0^2)-i \om \g^+(\om) }~  \label{gom}
\eea
with $\g^+(\om)=\g^+_R(\om)+ i \g^+_I (\om)$ where the real and imaginary
parts are given by:
$\g^+_R(\om)= \sum_\a ({\pi c_\a^2}/{2 \om_\a^2})
  [\d(\om+\om_\a)+\d(\om-\om_\a)],~~ 
\g^+_I(\om) =  \sum_\a {c_\a^2 \om}/{[ \om_\a^2 (\om^2-\om_\a^2)]}$.  
Now from eq.~(\ref{gom}) it follows that  $G^+(t \to \infty)=
\lim_{t \to \infty} (1/2\pi) \int d\om  e^{-i \om t }
G^+(\om)$ will vanish (this follows from the Riemann-Lebesgue lemma
\cite{bender})  {\emph{unless}} $G^+(\om)$ blows up at some real
value of $\om$.
Thus the condition under which $G^+(t\to \infty)$ does not vanish  is
that the equation     
\bea
- (\om^2-\om_0^2)-i \om \g^+(\om)=0 \label{bseq}
\eea
has a real $\om$ solution, which we will denote by $\Om_b$. It follows from
eq.~(\ref{bseq}) that $\g_R(\Om_b)=0$ which means that $\Om_b$
necessarily lies outside the bath band-width. 
This solution corresponds to a bound state. To see this we now solve 
the equations of motion of the coupled system-plus-bath 
by a resolution into its normal modes. Let us denote the normal mode
frequencies by $\Om_Q$ and the corresponding eigenfunction by
${\bf{U}}_Q=\{ U_{0,Q},U_{\a=1,Q},...U_{\a=N,Q}\}$ where $U_{0Q}$
corresponds to the system variable $x$. They satisfy the
equations:
\bea
- \Om_Q^2 U_{0Q}&=& - \om_0^2 U_{0Q} +\sum_\a c_\a \left( U_{\a Q} -\f{c_\a
U_{0Q}}{\om_\a^2} \right)\nn \\
-\Om_Q^2 U_{\a Q} &=& -\om_\a^2 \left(U_{\a Q}-\f{c_\a
  U_{0Q}}{\om_\a^2} \right)~ ~~\a=1...N~ \label{nmodeeq}
\eea   
A bound state \cite{ziman} occurs if there is a mode such that its
frequency $\Om_b$ 
lies outside the bandwidth of the isolated bath. The corresponding
eigenfunction ${\bf U}_b$ will have a finite weight at the system
point (\emph{i.e.} $U_{0b}= O(1)$). Solving, for $U_{\a b}$, the second
  equation in eq.~(\ref{nmodeeq}) and substituting into the first
  equation, we find that the condition for a bound 
state is given precisely by the solution (if it exists) of eq.~(\ref{bseq}).
The corresponding eigenfunction is given by:
$U_{\a b}={c_\a U_{0b}}/({-\Om_b^2+\om_\a^2}),~~\a=1...N$ and 
$U_{0 b} =  [ 1+ \sum_\a {c^2_\a}/{(-\Om_b^2+\om_\a^2)^2}]^{-1/2}$.
The general solution of the full equations of motion in terms of
normal modes is 
\bea
Y=U\cos{\Om (t-t_0)} U^{-1} Y(t_0) + U \f{\sin{
\Om(t-t_0)}}{ \Om} U^{-1} \dot{Y}(t_0) ~, \nn
\eea
where
$Y=\{x,X_1,...X_\a...X_N\}^T$. From this one can identify
the part of $x(t)$ involving 
$p(t_0)$. Comparing with eq.~(\ref{gensol}) we get 
$G(t)=\sum_Q U^2_{0Q} {\sin{ (\Om_Q t)}}/{\Om_Q}$.   
If there are no bound states then the frequencies $\Om_Q$ form a
continuous spectrum, the sum can be converted into an
integral, and in the limit $t \to \infty$ the infinite oscillations
lead to the integral vanishing. In the presence of a bound state we
get a non-vanishing contribution given by 
\bea
G(t \to \infty) = U_{0b}^2 \f{\sin{( \Om_b t )}}{\Om_b}~. \label{Ginf}
\eea 
Thus we again arrive at the conclusion that in the presence of bound
states, thermal equilibration is not achieved and the steady state
properties depend on the initial conditions of the system.
The long time form in eq.~(\ref{Ginf})  can also be obtained directly
from eq.~(\ref{gom}) by integrating 
around the singularity given by eq.~(\ref{bseq}).  
In the absence of bound states, eq.~\ref{gensol} gives the  unique
steady state solution $x(t)=\int_{-\infty}^\infty dt' G^+(t-t')
\n(t')$, and one can verify that:
\bea
\la \f{\om_0^2 x^2}{2} \ra = \f{\om_0^2 k_B T}{2 \pi} 
\int_{-\infty}^\infty d \om |G^+(\om)|^2 Re[\g^+ (\om)] \nn &=&\f{k_B T}{2}\\ 
\la \f{p^2}{2} \ra = \f{ k_B T}{2 \pi} 
\int_{-\infty}^\infty d \om~ \om^2 |G^+(\om)|^2 Re[\g^+ (\om)] 
&=&\f{k_B T}{2} ~,\nn
\eea
where the integrals are easy to evaluate if one chooses appropriate
contours \cite{kubo}. The inset in fig.~(\ref{equilib}) gives a
numerical demonstration of the nonequilibration problem (see following
section for details of numerics). 

\section{ Equilibration in a nonlinear potential}
Usually one
expects that introducing nonlinearity should help in equilibration and
we will now test this expectation. We will consider a particular model
of a heat bath for which it will be  possible to analyze things in
detail. This is the Rubin model in which the bath  is a one-dimensional
chain of coupled oscillators. Thus the full Hamiltonian is:
$H=H_s+\sum_{i=1}^N [ {p_i^2}/{2} +{(x_{i+1}-x_{i})^2}/{2}
  ] + {(x_1-x)^2}/{2}$,  
with $x_{N+1}=0$ and where $x=x_0$ refers to the system. 
Transforming to normal mode coordinates of the bath we recover the
form eq.~(\ref{ham}) with $\om_\a=4 \sin^2(q_\a/2)$ and $c_\a = 
[2/(N+1)]^{1/2} \sin (q_\a)$ with $q_\a=\a \pi/(N+1)$ and $\a=1,2...N$. 
From this we can find the explicit form of $\g^+(\om)$ which is:
\bea
&& i\om \g^+(\om)=-(1-e^{iq}) ~~~{\rm for}~|\om|=2| \sin{(q/2)}| ~<~2 \nn
\\
&&=-(1+e^{-\s}) ~~~{\rm for}~|\om|=2 \cosh{(\s/2)} ~>~2.   \label{alpeq} 
\eea
For a harmonic potential $V(x)=\om_0^2 x^2/2$ we see from
eq.~(\ref{bseq})  that the condition
for getting bound states is $ \om_0^2 > 2 $. The bound-state mode
is given by $x_i^b=A  (-1)^i e^{-\s i}\cos(\Om_b t)$.
where 
$\Om_b=\om_0^2/\sqrt{\om_0^2-1}$ is the bound state frequency,
$\s=\ln{(\om^2_0-1)}$ and $A$ is an arbitrary 
constant which can be fixed by normalization.    

\begin{figure}
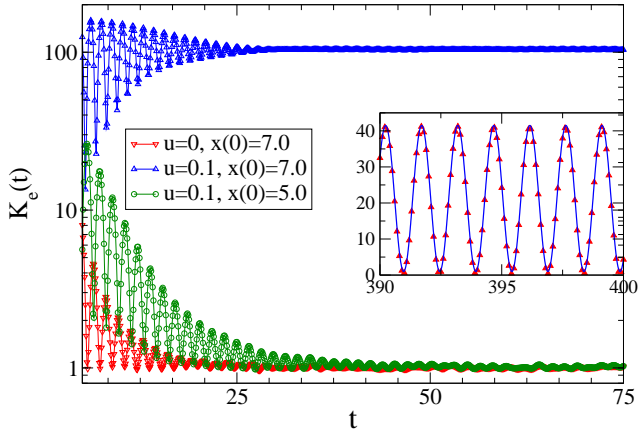

\onefigure[width=3.3in]{fig1p.eps}
\caption{Plot of average kinetic energy $K_e(t)$ as a function of time
  for different initial conditions. The purely linear case always
  equilibrates (for $\om_o=1$) while in the nonlinear case, equilibration  depends on
  initial conditions. The behaviour of $P_e(t)$ is similar. The inset shows the
nonequilibration problem in the linear case. The parameter
value $\om_0^2=3.0$ gives a bound state with $\Om_b=3/2^{1/2}$ and we
plot $K_e(t)$ at large times for $x(0)=5.0$ and
$T=1$. The solid line gives the analytic prediction $K_e(t)=\ddot
{G}(t)^2 x(0)^2 +T$ with $G(t)$ given by eq.~(\ref{Ginf}).   
}
\label{equilib}
\end{figure}

We now introduce the nonlinear term $u x^4/4$ and study its effect
numerically. Let us choose $\om_0^2=1$, in
which case, in the absence of the nonlinear term, there are no bound
states and we expect equilibration. In our simulation we consider a
heat bath with $N=1000$ particles whose initial positions and momenta
are chosen from the  Boltzmann distribution $e^{- H_b/T}$ with
$T=1$.  
The system has the initial state $\{p(0)=0, x(0)\}$. We
solve the equations of motion numerically with this initial state with
the system-bath coupling turned on at time $t=0$. We compute the expectation
values $K_e(t)=\la p^2 \ra$ and $P_e(t)=\la x ~\p H/\p x \ra$ where
$\la....\ra$ denotes an average over the bath initial conditions. In
our simulations we averaged over $10^4$ initial conditions. We have
checked that results do not change on  increasing the bath size or
number of aveages. For
equilibration we expect the steady state values $K_e=P_e=T$.    
In fig.~\ref{equilib} we see that for the initial condition $x(0)=7.0$
the linear problem equilibrates as expected. On the other hand for a
small value of the nonlinearity parameter $u=0.1$ the system \emph{fails to 
  equilibrate}. For a different initial condition $x(0)=5.0$ the system
equilibrates. 
The loss of equilibration infact occurs whenever $x(0)$ is greater than
a critical value $x_c\approx 5.6$. 

\begin{figure}
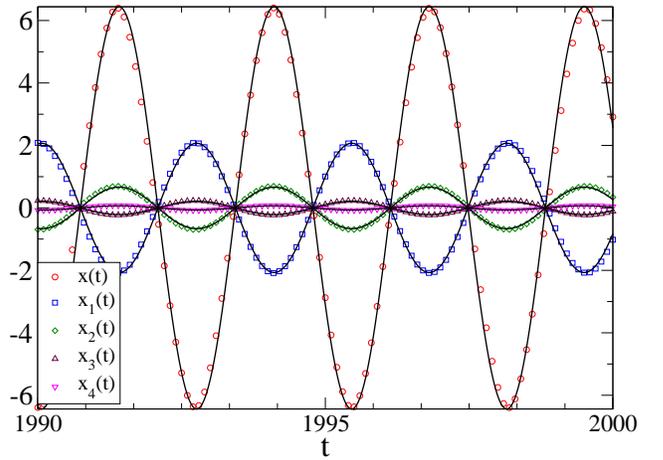

\onefigure[width=3.3in]{fig2.eps}
\caption{Plots of the positions of various particles $x(t)$ and
  $x_i(t)$ for $i=1,2,3,4$ as functions of time plotted after the
  system has reached a steady state. Initial condition was $x(0)=7.0$
  and all other positions and momenta set to zero. The solid lines
  correspond to the analytical prediction for  the breather mode with 
  $A=6.44$, $\s=1.135$ and $\Om_b=2.331$. }
\label{breath} 
\end{figure}

The dependence of equilibration on initial
conditions can be traced to the formation of localized states that can 
be generated once we have nonlinearity.  
These {\emph{nonlinear localized modes}} \cite{sievers,flach} are stable time-dependent
solutions of the equations of motion that are localized and largely
monochromatic and have the same form as the impurity bound states
namely  $x^b_i(t)=A (-1)^i e^{-\s i}
\cos(\Om_b t)$. If we plug this into the equations of motion and
neglect terms containing higher harmonics ($\sim 3 \Om_b$) 
we then get the following
conditions relating $\Om_b, \s$ and $A$ (for the parameter choice $\om_0=1$):
\bea
A^2=\f{4 e^\s}{3 u},~~~~
\Om^2_b=2+e^{\s}+e^{-\s}~. 
\label{bscond}
\eea 
Note that, of the three parameters needed to describe a bound state
namely $\Om_b, \s, A$ only two get fixed and we thus have a continuum
set of possible bound states which can be described by a single
parameter.  
It is clear that for any non-zero $u$ we can always choose 
$\s >0$ and determine $A,~\Om_b$ from eq.~(\ref{bscond}). Hence bound
states exist for \emph{any} value of $u$. 
A clean  way to numerically observe these
localized modes is by initially preparing the bath at $T=0$ and
the system with initial conditions $\{p(0)=0, x(0)\}$. Then at very
long times we find that this initial condition relaxes to a localized
mode  if $x(0)$ is sufficiently large. 
In fig.~\ref{breath} we plot the coordinates $\{x(t),x_i(t)\}$ for
$i=1,2,3,4$ as functions of time (for parameter values $\om_0=1$
and $u=0.1$) . For the initial 
condition $x(0)=7.0$  it is clear that the long time solution is
quite accurately described by the localized mode solution. 
In the fits with the localized mode we fix $A$ from the numerically
obtained amplitude of $x(t)$  and obtain  $\s$ and $\Om_b$ from
eq.~(\ref{bscond}). 
For the  initial condition $x(0)=5.0$ we find that the initial energy
quickly dissipates into the bath and at long times is distributed uniformly.
A more accurate form for the localized mode can be obtained by
including higher harmonics. At the next order, the solution is given by 
$x_i^b=A^{(1)}(-1)^ie^{-\s^{(1)} i} \cos{\Om_b t}+
A^{(2)}(-1)^ie^{-\s^{(2)} i} \cos{3\Om_b t}$, where again there is only
one unknown constant. In our example we find that $A^{(2)} << A^{(1)}$
(For all values of $u$ we get $A^{(2)}/A^{(1)} < 1/21$). 

Thus we have shown that the finite-temperature non-equilibration
problem is related to the formation
of nonlinear localized modes which, unlike in the harmonic case, 
depends on initial conditions.  

\section{Discussion}
In summary we have examined  the  conditions
necessary for thermal equilibration of a  particle in a  
potential well and evolving through a generalized Langevin equation.
For the case of a harmonic potential we show that 
for finite band-width baths the system will not always equilibrate. 
Using the microscopic model of a bath as a collection of oscillators
we show that non-equilibration arises because of the formation of
bound states in the coupled system-plus-bath. 
Surprisingly we find that making the system nonlinear does not
restore equilibration. On the contrary nonlinearity assists in the
formation of localized modes and hence causes loss of equilibration for
arbitrary  initial conditions. These nonlinear localized modes
are usually studied in the context of periodic nonlinear lattices,
while here they arise in a situation where only one spring is nonlinear. 
In this paper we have considered a special one-dimensional
bath. However localized modes also occur in higher dimensional
nonlinear lattices with finite band-widhts \cite{flach} and hence we
expect that the problem of equilibration is quite general.

In Ref.~\cite{dhar}  electronic systems described by the 
tight binding noninteracting Hamiltonian were discussed and it was 
shown that the formation of bound states with finite band width
reservoirs leads to a similar problem of equilibration. One example
which was studied was that of a single site attached to a
one-dimensional free electron reservoir. 
An interesting question that the present study raises is whether
the introduction of an onsite interaction, say of the  Hubbard type,  
would lead to similar effects as the nonlinear term in the case of the
oscillator.

\end{document}